\documentclass[twocolumn,prb,superbib,superscriptaddress,showpacs]{revtex4}
\usepackage{amsfonts}
\usepackage{amsmath}
\usepackage{amssymb}
\usepackage{graphicx}%
\providecommand{\U}[1]{\protect\rule{.1in}{.1in}}

\newcommand{\eff}{\text{eff}}
\newcommand{\CW}{\text{CW}}
\newcommand{\mg}{\text{mag}}
\newcommand{\ph}{\text{phon}}
\newcommand{\AFM}{\text{AFM}}
\newcommand{\FM}{\text{FM}}
\newcommand{\Iv}{\mathbf{I}}
\newcommand{\Sv}{\mathbf{S}}

\begin{document}
\title{Strong frustration due to competing ferromagnetic and antiferromagnetic interactions: magnetic properties of M(VO)$_{2}$(PO$_{4}$)$_{2}$ (M = Ca and Sr)}
\author{R. Nath}
\email{ramesh_phy2003@yahoo.com} \affiliation{Max-Planck-Institut
f\"{u}r Chemische Physik fester Stoffe, N\"{o}thnitzer Str. 40,
01187 Dresden, Germany}
\author{A.~A.~Tsirlin}
\email{altsirlin@gmail.com} \affiliation{Max-Planck-Institut f\"{u}r
Chemische Physik fester Stoffe, N\"{o}thnitzer Str. 40, 01187
Dresden, Germany} \affiliation{Department of Chemistry, Moscow State
University, 119992 Moscow, Russia}
\author{E.~E.~Kaul}
\altaffiliation[Present address: ]{Low Temperature Group, Bariloche
Atomic Centre -- National Commission of Atomic Energy. Av. Bustillo
9500 (C. P. 8400) S. C. de Bariloche, Argentine.}
\affiliation{Max-Planck-Institut f\"{u}r Chemische Physik fester
Stoffe, N\"{o}thnitzer Str. 40, 01187 Dresden, Germany}
\author{M.~Baenitz}
\affiliation{Max-Planck-Institut f\"{u}r Chemische Physik fester
Stoffe, N\"{o}thnitzer Str. 40, 01187 Dresden, Germany}
\author{N.~B\"{u}ttgen}
\affiliation{Experimentalphysik V, Elektronische Korrelationen und
Magnetismus, University of Augsburg, D-86135 Augsburg, Germany}
\author{C. Geibel}
\affiliation{Max-Planck-Institut f\"{u}r Chemische Physik fester
Stoffe, N\"{o}thnitzer Str. 40, 01187 Dresden, Germany}
\author{H. Rosner}
\affiliation{Max-Planck-Institut f\"{u}r Chemische Physik fester
Stoffe, N\"{o}thnitzer Str. 40, 01187 Dresden, Germany}

\begin{abstract}
We present a detailed investigation of the magnetic properties of
complex vanadium phosphates M(VO)$_{2}$(PO$_{4}$)$_{2}$ (M = Ca and Sr)
by means of magnetization, specific heat, $^{31}$P NMR measurements,
and band structure calculations. Experimental data evidence the
presence of ferromagnetic and antiferromagnetic interactions in
M(VO)$_{2}$(PO$_{4}$)$_{2}$, resulting in a nearly vanishing
Curie-Weiss temperature \mbox{$\theta_{\CW}\leq 1$~K} that contrasts with
the maximum of magnetic susceptibility at 3 K. Specific heat and NMR
measurements also reveal weak exchange couplings with the thermodynamic
energy scale $J_c=10-15$~K. Additionally, the reduced maximum of the
magnetic specific heat indicates strong frustration of the spin
system. Band structure calculations show that the spin systems of
the M(VO)$_{2}$(PO$_{4}$)$_{2}$ compounds are essentially
three-dimensional with the frustration caused by competing ferromagnetic
and antiferromagnetic interactions. Both calcium and strontium
compounds undergo antiferromagnetic long-range ordering at $T_N=1.5$
K and 1.9 K, respectively. The spin model reveals an unusual example
of controllable frustration in three-dimensional magnetic systems.
\end{abstract}

\maketitle

\section{Introduction}

Low-dimensional and/or frustrated spin systems are a subject of
thorough studies due to their variety of unusual ground states and quantum phenomena.\cite{anderson1987,park2007} One of the most striking
effects in these systems is the formation of a spin-liquid, a
dynamically disordered ground state with short-range magnetic
correlations. Spin liquids are caused by quantum fluctuations that impede or even suppress long-range magnetic ordering

Quantum fluctuations are enhanced in systems with low spin value, low dimensionality, and/or magnetic frustration. At present, numerous spin models (and the respective structural types) for the frustrated spin systems are known.\cite{greedan2001,harrison2004} Most of these systems are geometrically frustrated magnets, since the frustration is caused by the topology of the lattice. However, several models reveal frustration due to a specific topology of magnetic interactions, while the lattice itself does not prevent the system from long-range spin ordering. An important advantage of the latter systems is the possibility to vary the degree of frustration by tuning individual exchange couplings. Hence, the formation of different ground states within one model and the access to quantum critical points are possible. Below, we focus on the respective models and briefly review both one-dimensional (1D) and two-dimensional (2D) cases.

The simplest 1D lattice is a uniform chain. To frustrate such a
system, one should introduce next-nearest-neighbor (NNN) interaction
$J_2$. Typically, $J_2$ is antiferromagnetic and competes with
either ferromagnetic (FM) or antiferromagnetic (AFM)
nearest-neighbor (NN) interactions $J_1$. The resulting physics
strongly depends on the sign of $J_1$. Thus, for $J_1>0$ (i.e., for
AFM NN coupling) the system has a dimerized ground state for
$J_2/J_1>0.241$ that may be further stabilized by lattice distortion
as experimentally observed in the spin-Peierls compound
CuGeO$_3$.\cite{hase1993,castilla1995} If $J_1<0$, helical magnetic
ordering is formed for $J_2/J_1<-0.25$, and the pitch angle of the
spiral depends on the $J_2/J_1$ ratio. Recently, a number of
frustrated chain systems with $J_2/J_1<0$ have been
studied,\cite{drechsler2007-r} and even the vicinity of the quantum
critical point at $J_2/J_1=-0.25$ was accessed
experimentally.\cite{drechsler2007} In these systems, the
possibility to tune the pitch angle by varying the exchange
couplings may be particularly attractive due to another intriguing
property -- the magnetoelectric effect observed in LiCu$_2$O$_2$ and
LiCuVO$_4$.\cite{park2007,naito2007,schrettle2008,yasui2008,seki2008}

In two dimensions, we will consider the square lattice. Frustration
can be introduced by diagonal NNN interactions ($J_2$) that compete
with NN interactions $J_1$ running along the side of the square. The
resulting model is known as a frustrated square lattice (FSL) or
$J_1-J_2$ model. The phase diagram of the FSL has been extensively
studied theoretically (see, e.g., Refs.
\onlinecite{chandra1988,sushkov2001,siurakshina2001,shannon2004}).
It shows three ordered phases (ferromagnet, N\'eel antiferromagnet,
and columnar antiferromagnet) as well as two critical regions near
the quantum critical points at $J_2/J_1=\pm 0.5$. Basically, these
regions are supposed to reveal spin-liquid ground states, although
there is also a proposition of a nematic order at
$J_2/J_1=-0.5$.\cite{shannon2006} Yet both the conjectures are not
verified experimentally, since most of the FSL systems studied so
far show columnar magnetic ordering and do not fall to the critical
regions. Nevertheless, the detailed investigation of the respective
compounds was highly helpful and disclosed important methodological
aspects of studying the frustrated spin systems and understanding
their physics.\cite{melzi2000,melzi2001,rosner2002,bombardi2004}

Based on the above representative examples, one may expect that
similar changes of the ground state due to the alteration of the
exchange couplings could be realized in other frustrated models --
either in one, two, or even three dimensions. This consideration and
the recent investigation of an interesting FSL system, the complex
vanadium phosphate Pb$_2$VO(PO$_4)_2$ (Ref. \onlinecite{kaul2004}),
motivated us to start a systematic research of the related
compounds. Below, we present the study of M(VO)$_2$(PO$_4)_2$ (M =
Ca, Sr) -- two representatives of the same compound family that
reveal three-dimensional (3D) spin systems with the frustration
caused by the specific topology and magnitudes of the exchange
interactions. In the following, we show clear evidence of the
frustration in the results of several experimental techniques:
magnetization, specific heat, and nuclear magnetic resonance (NMR)
measurements. In order to understand the microscopic origin of the
frustration, we use band structure calculations.

The outline of the paper is as follows. In Sec.~\ref{structure}, we analyze the crystal structure of the M(VO)$_{2}$(PO$_{4}$)$_{2}$ compounds focusing on possible paths of the exchange interactions. Methodological aspects are reviewed in Sec.~\ref{method}. Section~\ref{experiment} presents experimental results including magnetization, specific heat, and $^{31}$P NMR. In Sec.~\ref{band}, we report the results of band structure calculations and construct a microscopic model of the exchange interactions in Ca(VO)$_{2}$(PO$_{4}$)$_{2}$. Experimental and theoretical results are compared and discussed in Sec.~\ref{discussion} followed by our conclusions.

\section{Crystal structure}
\label{structure}
\begin{figure}
\includegraphics{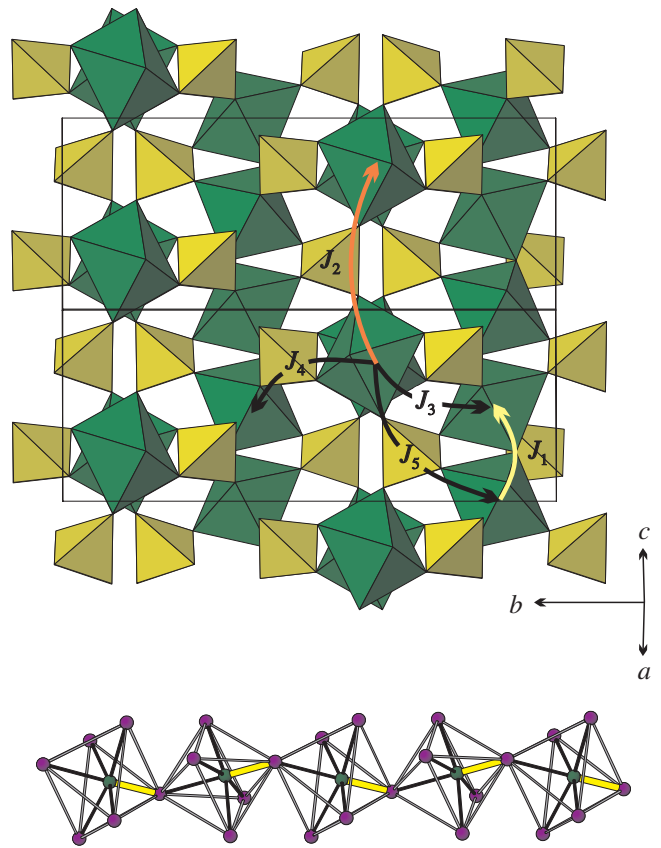}
\caption{\label{fig_struct}
(Color online) Projection of the Ca(VO)$_{2}$(PO$_{4}$)$_{2}$ crystal structure along the [101] direction (upper panel) and the chain of corner-sharing VO$_{6}$ octahedra (bottom panel). In the upper panel, the arrows denote different paths of superexchange interactions running via PO$_4$ tetrahedra. To simplify the figure, calcium atoms are not shown. In the bottom panel, thick (yellow) lines indicate short V--O bonds that are aligned along the direction of the chain.}
\end{figure}

Complex vanadium phosphates Ca(VO)$_{2}$(PO$_{4}$)$_{2}$ (Ref.~\onlinecite{lii1992}) and Sr(VO)$_{2}$(PO$_{4}$)$_{2}$ (Ref.~\onlinecite{berrah1999}) are isostructural and crystallize in a face-centered orthorhombic unit cell (space group $Fdd2$, $Z=8$). The structure presents one vanadium site that shows distorted octahedral coordination. The V$^{+4}$O$_{6}$ octahedra share corners to form
chains running along the $[101]$ and $[10\bar{1}]$ directions. The chains are
joined into a 3D-framework by PO$_{4}$ tetrahedra (Fig.~\ref{fig_struct}).

Magnetic interactions in insulating transition-metal compounds
mainly depend on two factors: the orbital state of the magnetic
cation and the possible paths for superexchange interactions. In the
following, we will discuss both of these factors.
\begin{figure*}
\includegraphics{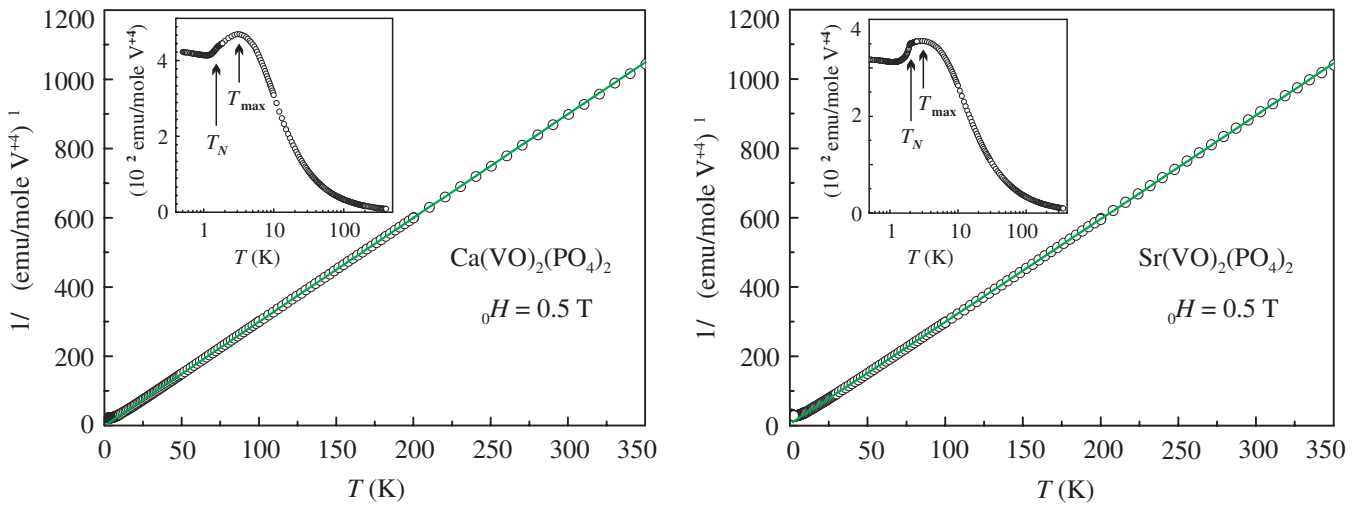}
\caption{\label{fig_chi} (Color online) Reciprocal magnetic
susceptibility ($1/\chi$) vs. temperature ($T$) measured at $\mu
_{0}H=0.5$ T down to $0.4$ K for Ca(VO)$_{2}$(PO$_{4}$)$_{2}$ (left
panel) and Sr(VO)$_{2}$(PO$_{4}$)$_{2}$ (right panel). The solid
lines are best fits of the data to Eq. (\ref{Curie}). In the insets,
$\chi$ vs. $T$ is plotted in the logarithmic scale to focus on the
$T_{\max}^{\chi}$ and $T_{N}$.}
\end{figure*}

The orbital state of the magnetic cation is determined by the local environment. In the M(VO)$_{2}$(PO$_{4}$)$_{2}$ compounds, vanadium has distorted octahedral coordination with one short bond (see bottom part of Fig.~\ref{fig_struct}). This type of environment is typical for V$^{+4}$ (electronic configuration $3d^1$) and gives rise to a non-degenerate ground state with half-filled $d_{xy}$ orbital (see Refs. \onlinecite{tsirlin2008} and \onlinecite{kaul2003} and Sec.~\ref{band}). The $d_{xy}$ orbital is located in the plane perpendicular to the short V--O bond. Short bonds are aligned along the structural chains, therefore the overlap of the half-filled orbital with $p$ orbitals of the two bridging oxygen atoms is very weak, and V--O--V is not an efficient path for superexchange interactions. On the other hand, the $d_{xy}$ orbital may overlap with the $p$ orbitals of four other oxygen atoms providing more complicated V--O--P--O--V paths for the superexchange.

The chains of the VO$_{6}$ octahedra in the structures of the M(VO)$_{2}$(PO$_{4}$)$_{2}$ compounds are not parallel, and the V--P--O framework looks quite complicated. In order to simplify the situation, one should take into account two considerations: (i) Every vanadium atom is connected to four different PO$_{4}$ tetrahedra. (ii) Every PO$_{4}$ tetrahedron is connected to four different vanadium atoms. Thus, every vanadium atom may have $3\times4=12$ different V--O--P--O--V connections at the most. In fact, we find only ten of them, since two connections (namely, that corresponding to $J_{3}$) are provided by two tetrahedra simultaneously (see Fig.~\ref{fig_struct}). Some of the connections are equivalent by symmetry and hence give rise to identical exchange interactions. Finally, we find that every vanadium atom can take part in five different superexchange interactions running via PO$_4$ tetrahedra, and there are two interactions of each type per vanadium atom. We denote these interactions and corresponding hoppings by $J_{1}-J_{5}$ and $t_{1}-t_{5}$, respectively (see Fig.~\ref{fig_struct}). $J_{1}$ runs along structural chains, $J_{2}$ runs between parallel structural chains, while $J_{3}-J_{5}$ provide couplings between non-parallel chains.

According to a recent study of another vanadium phosphate,
Ag$_2$VOP$_2$O$_7$ (Ref.~\onlinecite{tsirlin2008}), exchange
couplings via PO$_4$ tetrahedra can hardly be estimated using
structural data only. Such couplings depend on numerous geometrical
parameters, and slight structural alterations may give rise to a
huge change of the exchange integrals. Therefore, we do not attempt
to construct a spin model based on qualitative and empirical
considerations. Instead, we proceed to the results of band structure
calculations and estimate individual exchange couplings
(Sec.~\ref{band}). One should be aware that the spin systems of the
M(VO)$_2$(PO$_4)_2$ compounds are intricate and quite complicated
even for the computational study due to the weakness of the exchange
couplings and the presence of numerous NN and NNN exchange pathways.
Therefore, the computational estimates are less accurate as compared
to that in other spin-1/2 systems (see, e.g., Refs.
\onlinecite{drechsler2007}, \onlinecite{rosner2002}, and
\onlinecite{tsirlin2008}). Nevertheless, we succeed to reveal the
leading exchange couplings and suggest a plausible scenario of the
magnetic frustration in the M(VO)$_2$(PO$_4)_2$ compounds (see
Sec.~\ref{discussion}).

\section{Methods}
\label{method} Polycrystalline samples of
M(VO)$_{2}$(PO$_{4}$)$_{2}$ were prepared by solid state reaction
technique using CaCO$_{3}$ ($99.9$\%), SrCO$_{3}$ ($99.999$\%),
VO$_{2}$ ($99.99$\%), and (NH$_{4})_{2}$HPO$_{4}$ ($99.9$\%) as
starting materials. The synthesis involved two steps. In the first
step, the intermediate compounds MP$_{2}$O$_{6}$ were prepared by
firing the stoichiometric mixtures of MCO$_3$ and (NH$_4)_2$HPO$_4$
at $850$ $^{\circ}$C for 1 day in air with one intermediate
grinding. In the second step, the intermediate products were mixed
with VO$_{2}$ in the appropriate molar ratio and heated for 4
days in dynamic vacuum (10$^{-5}$ mbar) with several intermediate
grindings and pelletizations. The annealing temperatures were 900
and 850 $^{\circ}$C for Ca(VO)$_{2}$(PO$_{4}$)$_{2}$ and
Sr(VO)$_{2}$(PO$_{4}$)$_{2}$, respectively. The resulting samples
were single-phase as confirmed by X-ray diffraction (STOE powder
diffractometer, CuK$_{\alpha}$ radiation). Lattice parameters were
calculated using a least-squares fit procedure. The obtained lattice
parameters for Ca(VO)$_{2}$(PO$_{4}$)$_{2}$ [$a=11.774(2)$ \AA ,
$b=15.777(3)$ \AA , and $c=7.163(1)$ \AA] and
Sr(VO)$_{2}$(PO$_{4}$)$_{2}$ [$a=12.008(2)$ \AA , $b=15.924(3)$ \AA
, and $c=7.207(1)$ \AA] are in agreement with the previously
reported values.\cite{lii1992, berrah1999}

Magnetization data were measured as a function of temperature
and field using a SQUID magnetometer (Quantum Design MPMS). Heat
capacity data were collected with Quantum Design PPMS on
pressed pellets using the relaxation technique. All the measurements
were carried out over a large temperature range ($0.4$ K $\leq$ $T$
$\leq$ $400$ K) in magnetic fields $\mu _{0}H$ up to 14 T. The
low-temperature measurements were done partly using an additional
$^{3}$He setup.

The NMR measurements were carried out using pulsed NMR techniques on
$^{31}$P (nuclear spin $I=1/2$ and gyromagnetic ratio $\gamma/2\pi$
= $17.237$ MHz/Tesla) nuclei in a large temperature range ($1$ K
$\leq$ $T$ $\leq$ $300$ K). We have done the measurements at a radio
frequency of $70$~MHz which corresponds to an applied field of about
$4.06$ T. Low-temperature NMR measurements were partly done using a
$^{3}$He/$^{4}$He dilution refrigerator (Oxford Instruments) with
the resonant circuit inside the mixing chamber. Spectra were
obtained either by Fourier transform (FT) of the NMR echo signals or
by sweeping the field at fixed frequency of 70~MHz. The NMR shift
$K(T)=\left[ \nu\left( T\right)  -\nu_{\text{ref}}\right]
/\nu_{\text{ref}}$ was determined by measuring the resonance
frequency of the sample ($\nu\left(  T\right)  $) with respect to a
standard H$_{3}$PO$_{4}$ solution (resonance frequency
$\nu_{\text{ref}}$). The $^{31}$P spin-lattice relaxation rate
$(1/T_{1})$ was measured by using either a 180$^{\circ}$ pulse
(inversion recovery) or a comb of saturation pulses.

Scalar-relativistic band structure calculations were performed using the
full-potential local-orbital scheme (FPLO7.00-27)\cite{koepernik1999} and the parametrization of Perdew and Wang for the exchange and correlation potential.\cite{perdew1992} A $k$-mesh of 1728 points within the first Brillouin zone (510 in the irreducible part for the space group $Fdd2$) was used. Convergence with respect to the $k$-mesh was carefully checked.

First, a local density approximation (LDA) calculation was done using the full symmetry of the crystal structure (space group $Fdd2$). The results of this calculation enabled the choice of the orbital states relevant for the magnetic interactions. The respective bands were analyzed using a tight-binding (TB) model, and the resulting hoppings were applied to estimate the antiferromagnetic contributions to the exchange integrals. Such an analysis accounts for all reasonable paths of superexchange and provides a reliable basis for the microscopic model of the magnetic interactions. However, LDA does not give any information about ferromagnetic couplings in the system under investigation.

\begin{figure}[b]
\includegraphics{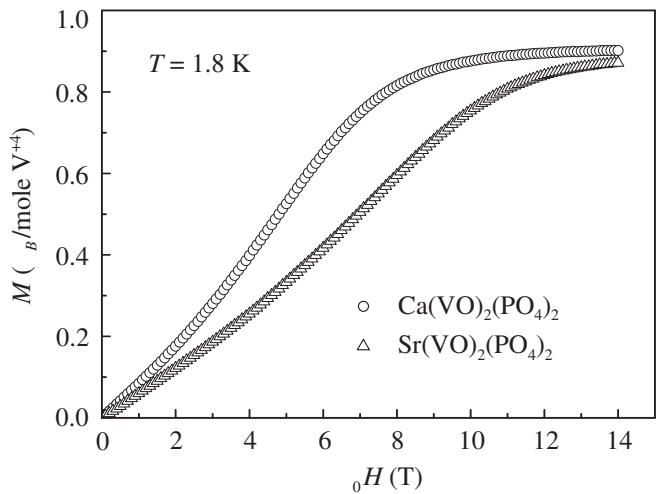}
\caption{\label{fig_mvsh}
Isothermal magnetization of the M(VO)$_2$(PO$_4)_2$ compounds measured at 1.8 K.}
\end{figure}
To account for ferromagnetic interactions in the microscopic model, we performed LSDA+$U$ calculations and compared total energies for four different patterns of spin ordering. Exchange couplings in the M(VO)$_2$(PO$_4)_2$ compounds are very weak; therefore one has to use the smallest unit cell (i.e., the primitive cell -- one-fourth of the face-centered unit cell) in order to achieve the highest possible accuracy. The primitive cell is triclinic (space group $P1$) and includes four inequivalent vanadium atoms, hence a number of different spin configurations can be formed.

\begin{table*}
\caption{The parameters of the Curie-Weiss fit [Eq. (\ref{Curie})] of the bulk susceptibility and the NMR shift data (marked by the superscripts $\chi$ and $K$, respectively). $\chi_0$ is the temperature-independent contribution, $\mu_{\eff}$ is the effective magnetic moment, and $\theta_{\CW}$ is the Curie-Weiss temperature. The listed standard deviations originate from the  least-squares fitting.}
\label{Chi_parameters}
\begin{ruledtabular}
\begin{tabular}{cccccc}\smallskip
Sample & $\chi_{0}$ ($10^{-6}$ emu/mole) & $\mu_{\eff}^{\chi}$
($\mu_B$) & $\theta_{\CW}^{\chi}$ (K) & $\mu_{\eff}^K$ ($\mu_B$) &
$\theta_{\CW}^{K}$ (K) \\\hline
Ca(VO)$_{2}$(PO$_{4}$)$_{2}$ & $2.1(5)$ & $1.630(1)$ & $-0.4(3)$ & $1.71(6)$ & $-0.3(1)$ \\
Sr(VO)$_{2}$(PO$_{4}$)$_{2}$ & $-1.5(5)$ & $1.653(1)$ & $1.0(1)$ & $1.70(5)$ & $-0.4(1)$ \\
\end{tabular}
\end{ruledtabular}\end{table*}
Following the results oin Ref.~\onlinecite{tsirlin2008}, we employed
several values of $U_d$ (Coulomb repulsion parameter of the LSDA+$U$
method) in our calculations. According to the previous
study,\cite{tsirlin2008} $U_d=6$ eV provides a reasonable
description of the octahedrally coordinated V$^{+4}$ within the FPLO
calculations. Yet one should keep in mind that the optimal value of
$U_d$ depends on numerous factors (basis set,\cite{foot5} local
environment of the transition-metal cation, and objective parameters),
and the unique choice of $U_d$ remains a subtle issue. Below, we use
several representative values of $U_d$ (4, 5, and 6 eV) and
carefully check their effect on the results. The exchange parameter
of LSDA+$U$ was fixed at $J=1$ eV since usually it has minor
importance compared to $U_d$, especially for the $3d^1$
configuration.

\section{Experimental results}
\label{experiment}
\subsection{Bulk susceptibility and magnetization}
Bulk magnetic susceptibilities ($\chi$) of the M(VO)$_2$(PO$_4)_2$
compounds are presented in Fig.~\ref{fig_chi}. With decreasing
temperature, the susceptibility increases in a Curie-Weiss manner
and passes through a broad maximum at $T_{\max}^{\chi}\simeq 3$ K.
The maximum is characteristic of low-dimensional spin systems and
indicates a crossover to a state with antiferromagnetic
correlations. The change in slope observed at $T_N=1.5$~K and $1.9$~K [for Ca(VO)$_{2}$(PO$_{4}$)$_{2}$ and
Sr(VO)$_{2}$(PO$_{4}$)$_{2}$, respectively] can be attributed to the
onset of long-range magnetic ordering.

To fit the bulk susceptibility data at high temperatures, we use the expression:
\begin{equation}
\chi=\chi_{0}+\frac{C}{T+\theta_{\CW}} \label{Curie}%
\end{equation}
where $\chi_{0}$ is the temperature-independent contribution that
accounts for core diamagnetism and Van Vleck paramagnetism, while
the second term is the Curie-Weiss law with the Curie constant
$C=N_A\mu_{\eff}^{2}/3k_B$. The data above 20 K were fitted with the
parameters listed in Table~\ref{Chi_parameters}. The resulting
effective moments $\mu_{\eff}^{\chi}$ are in reasonable agreement
with the spin-only value of $1.73\ \mu_B$, while the Curie-Weiss
temperatures $\theta_{\CW}^{\chi}$ for both the compounds are rather
small as compared to $T_{\max}^{\chi}$ (we use the superscript
$\chi$ to denote the values related to the analysis of the bulk
susceptibility data).

Basically, $T_{\max}^{\chi}$ characterizes the energy scale of the
magnetic interactions, and $\theta_{\CW}$ is a linear combination of
all the exchange integrals.\cite{johnston2000} Therefore, the
reduction in $\theta_{\CW}$ as compared to $T_{\max}^{\chi}$ implies
the presence of both FM and AFM interactions in the system under
investigation. Moreover, $T_N$ is considerably lower than
$T_{\max}^{\chi}$, and this effect may be caused by either low
dimensionality and/or frustration of the spin system. As we will
show below (Sec.~\ref{band}), the spin systems of the
M(VO)$_2$(PO$_4)_2$ compounds are three-dimensional. Thus, the
reduction in $T_N$ as compared to $T_{\max}^{\chi}$ indicates
magnetic frustration in M(VO)$_2$(PO$_4)_2$.

Field-dependent magnetization data for the M(VO)$_2$(PO$_4)_2$
compounds are shown in Fig.~\ref{fig_mvsh}. At low fields, the
curves reveal linear behavior, while a positive curvature is
observed at higher fields. Further increase in the field results in
the saturation at $\mu_0H_s\simeq 8$~T and $11.5$~T for
Ca(VO)$_2$(PO$_4)_2$ and Sr(VO)$_2$(PO$_4)_2$, respectively. The
saturation magnetization ($M_s$) is about $0.9\ \mu_B$/mole, i.e.,
slightly below the expected value of $1\ \mu_B$. The underestimate
of $M_s$ may be caused by non-magnetic impurities in the samples
under investigation. This explanation is further supported by the
slight reduction in $\mu_{\eff}$ and the magnetic entropy (see Sec.~\ref{heat}).

The saturation field shows the energy difference between
the ground state and the fully polarized (ferromagnetic) state of
the system. Thus, the values of $H_s$ can be used to estimate
exchange couplings, if the ground state of the system and the
leading exchange interactions are known. We will further discuss
this point in Sec.~\ref{discussion} and employ the $H_s$ values to
get quantitative information about exchange couplings in the systems
under investigation.
\begin{figure*}
\includegraphics{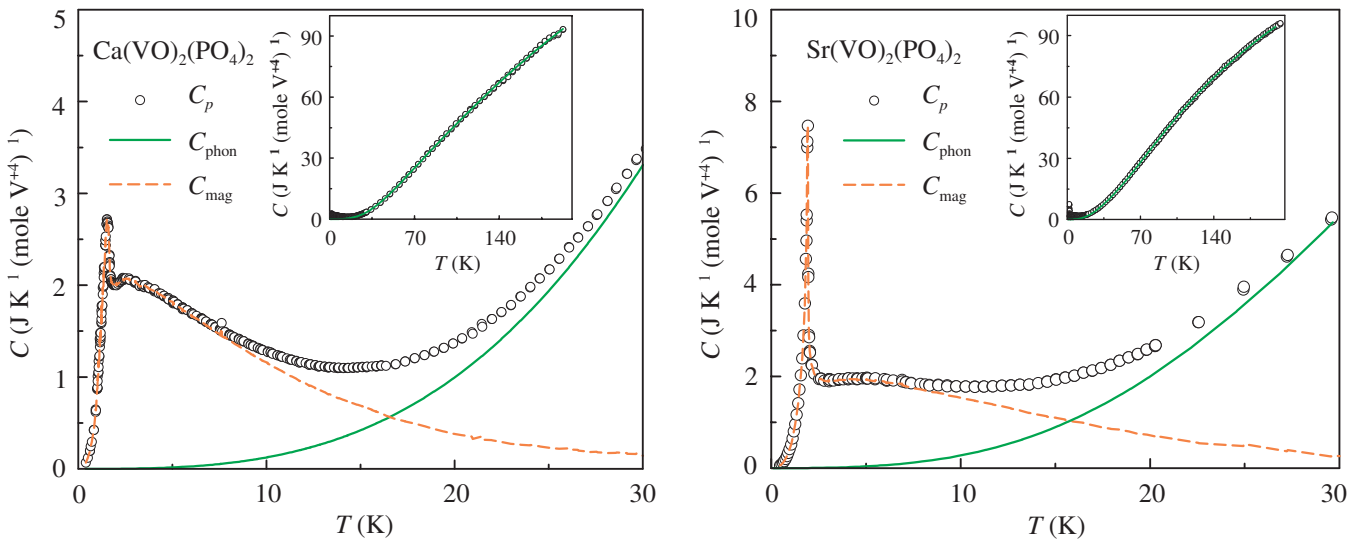}
\caption{\label{fig_cp} (Color online) Temperature dependence of the
specific heat measured at zero field for
Ca(VO)$_{2}$(PO$_{4}$)$_{2}$ (left panel) and
Sr(VO)$_{2}$(PO$_{4}$)$_{2}$ (right panel). The open circles are the
raw data, the solid lines show the phonon contribution $C_{\ph}$ as
found from the fit to Eq.~(\ref{Debye}), and the dashed lines denote
the magnetic contribution $C_{\mg}$. The insets show the fits with
Eq. (\ref{Debye}) in the wide temperature range.}
\end{figure*}

\subsection{Specific heat}
\label{heat} The specific heat ($C_p$) results at zero field are shown in
Fig.~\ref{fig_cp}. At high temperatures, $C_{p}$ is completely
dominated by the contribution of phonon excitations; therefore, both
the compounds reveal similar $C_p(T)$ curves. With decreasing
temperature, $C_{p}$ starts to increase below $12-15$ K indicating
that the magnetic part of the specific heat ($C_{\mg}$) becomes
prominent. With a further decrease in temperature, $C_{p}(T)$ shows a
broad maximum at around 3 K and 4 K due to the correlated spin
excitations and a sharp peak at $T_N=1.5$ K and 1.9 K associated
with the long-range magnetic ordering [the values are given for
Ca(VO)$_{2}$(PO$_{4}$)$_{2}$ and Sr(VO)$_{2}$(PO$_{4}$)$_{2}$,
respectively].\cite{foot7}

In order to get a quantitative estimate of $C_{\mg}$, the phonon
part $C_{\ph}$ was subtracted from the total measured specific heat
$C_{p}$. The general procedure is similar to that reported in
Refs.~\onlinecite{kini2006} and \onlinecite{nath2008}. The phonon part was estimated by fitting $C_{p}(T)$ at high temperatures \mbox{($15$~K $\leq T\leq 200$~K)} with a sum of Debye contributions. The
additional term $A/T^2$ accounted for the magnetic contribution, and
the final fit was performed using the equation:
\begin{equation}
C_{p}(T)=\frac{A}{T^{2}}+9R\sum_{n=1}^{n=4}c_{n}\left( \frac{T}{\theta_D^{(n)}} \right)^{3}\int\limits_{0}^{\theta_D^{(n)}/T}\frac{x^{4}e^{x}}{\left(  e^{x}-1\right)^2}dx \label{Debye}%
\end{equation}
where $R=8.314$ J/mole K is the gas constant, $\theta_D^{(n)}$ are the characteristic Debye temperatures, and $c_n$ are integer coefficients indicating the contributions of different atoms (or groups of atoms) to the specific heat. The phonon contribution was extrapolated down to
$0.4$ K and subtracted from the measured $C_{p}(T)$. The reliability of the whole procedure was justified by integrating $C_{\mg}/T$. The resulting magnetic entropies are 5.30 and 5.58 \mbox{J/mole K} [for Ca(VO)$_2$(PO$_4)_2$ and Sr(VO)$_2$(PO$_4)_2$, respectively] consistent with the expected value of $R\ln 2$.

High-temperature magnetic contribution $A/T^2$ is the lowest-order term in the high-temperature series expansion for the specific heat. According to Ref.~\onlinecite{johnston2000},
\begin{equation}
  A=\dfrac{3R}{32}\sum_i z_iJ_i^2
\end{equation}
where integers $z_i$ indicate the number of interactions $J_i$ for a single magnetic atom (i.e., coordination number for the interactions of type $J_i$). The structures of M(VO)$_2$(PO$_4)_2$ yield $z_i=2$ for any $i$ (see Sec.~\ref{structure}), and we find $A=(3R/16)J_c^2$ with the thermodynamic energy scale of the exchange couplings defined as $J_c=\sqrt{\sum_i J_i^2}$. Using the experimental values $A=142$ and 147 J K/mole, we estimate $J_c^{\,C}=9.6$ K and 9.8 K for the calcium and strontium compounds, respectively (the superscript $C$ denotes the values obtained from the analysis of the specific heat data).

\begin{figure}[b]
\includegraphics{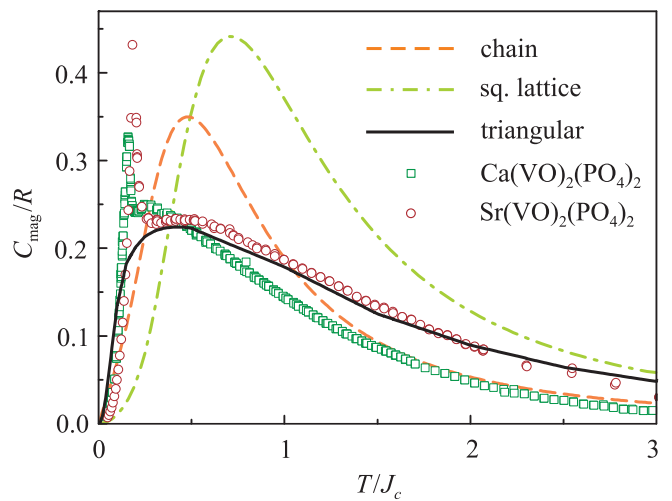}
\caption{\label{fig_cp-comparison}
(Color online) Magnetic contribution to the specific heat of M(VO)$_2$(PO$_4)_2$ and theoretical curves for the uniform chain (Refs.~\onlinecite{johnston2000} and \onlinecite{bernu2001}), non-frustrated square lattice (Refs.~\onlinecite{bernu2001} and \onlinecite{hofmann2003}), and triangular lattice (Ref.~\onlinecite{bernu2001}) models. The reduced temperature scale $T/J_c$ with $J_c=\sqrt{\sum_iJ_i^2}$ is used.}
\end{figure}

To interpret the specific heat data, we compare $C_{\mg}$ with the simulated curves for the representative spin-1/2 models in both one dimension (uniform chain) and two dimensions (non-frustrated square lattice, triangular lattice), see Fig.~\ref{fig_cp-comparison}. The maximum value of the magnetic specific heat ($C_{\mg}^{\,\max}$) and the shape of the maximum are characteristic of the magnitude of quantum fluctuations. The enhancement of quantum fluctuations suppresses correlated spin excitations, therefore $C_{\mg}^{\,\max}$ is reduced, and the maximum gets broader. One may follow this effect in Fig.~\ref{fig_cp-comparison}. The non-frustrated square lattice shows a high $C_{\mg}^{\,\max}\simeq 0.46R$.\cite{bernu2001,hofmann2003} In the 1D case, $C_{\mg}^{\,\max}$ is decreased to $0.35R$ due to stronger quantum fluctuations.\cite{johnston2000,bernu2001} The triangular lattice (a frustrated 2D system) shows an even lower ($C_{\mg}^{\,\max}=0.22R$) and broader maximum as compared to the uniform chain.\cite{bernu2001}

According to the results of band structure calculations
(Sec.~\ref{band}), the spin systems of M(VO)$_2$(PO$_4)_2$ are 3D.
The high dimensionality should result in a narrow maximum of the
magnetic specific heat with a high absolute value at the maximum.
However, the M(VO)$_2$(PO$_4)_2$ compounds reveal broad maxima with
the absolute values of about $0.25R$ and $0.23R$ for M = Ca and Sr,
respectively. Such absolute values are well below that for the
uniform chain (1D non-frustrated system) and nearly match the value
for the triangular lattice (2D frustrated system). This result
points to the presence of strong quantum fluctuations in the
M(VO)$_2$(PO$_4)_2$ compounds. The spin systems are 3D; therefore
the fluctuations should be entirely caused by the frustration. The
different breadth of the maxima for the calcium and strontium
compounds (see Fig.~\ref{fig_cp-comparison}) may also be an
indication of the stronger frustration in Sr(VO)$_2$(PO$_4)_2$.

\begin{figure}
\includegraphics{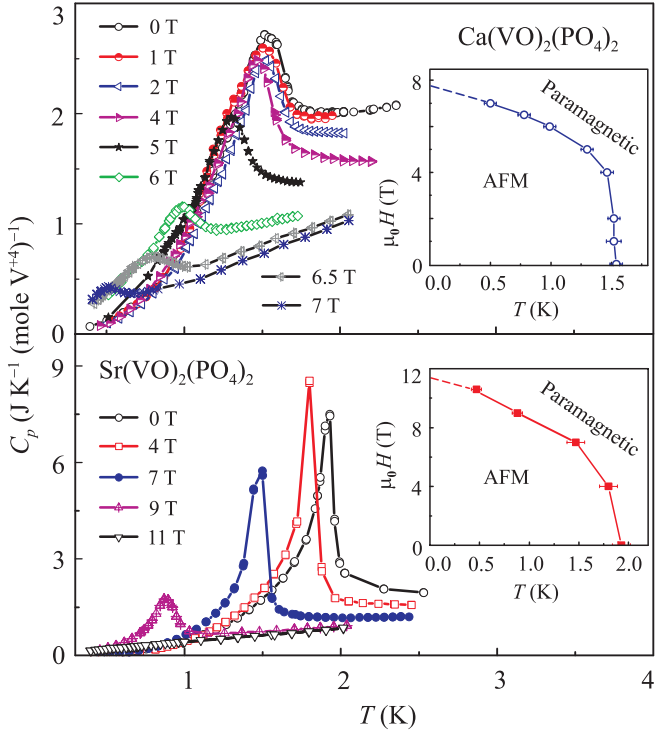}
\caption{\label{fig_cph}(Color online) Specific heat of Ca(VO)$_2$(PO$_4)_2$ (upper panel) and Sr(VO)$_2$(PO$_4)_4$ (bottom panel) measured at different applied fields ($H$). The insets show the respective $H-T$ phase diagrams.}
\end{figure}

The strong frustration evidenced by the specific heat data is
consistent with the considerable reduction in the ordering
temperature $T_N$ as compared to $T_{\max}^{\chi}$ (see the previous
subsection). However, in contrast to the geometrically frustrated
systems, $T_N$ does not vanish completely, and at sufficiently low
temperatures, long-range magnetic ordering is established. To
understand the nature of the ordered state, we studied field
dependence of the specific heat (Fig.~\ref{fig_cph}). At large, the
transition temperature gradually decreases with the increase in the
field and finally gets suppressed below 0.4 K at 8 and 11.5 T for
Ca(VO)$_2$(PO$_4)_2$ and Sr(VO)$_2$(PO$_4)_2$,
respectively.\cite{foot6} The $H-T$ phase diagrams shown in the
insets of Fig.~\ref{fig_cph} are typical for antiferromagnets and
point to antiferromagnetic ordering in the M(VO)$_2$(PO$_4)_2$
compounds.

\subsection{$^{31}$P NMR}
\label{NMR} 
\begin{figure}
\includegraphics{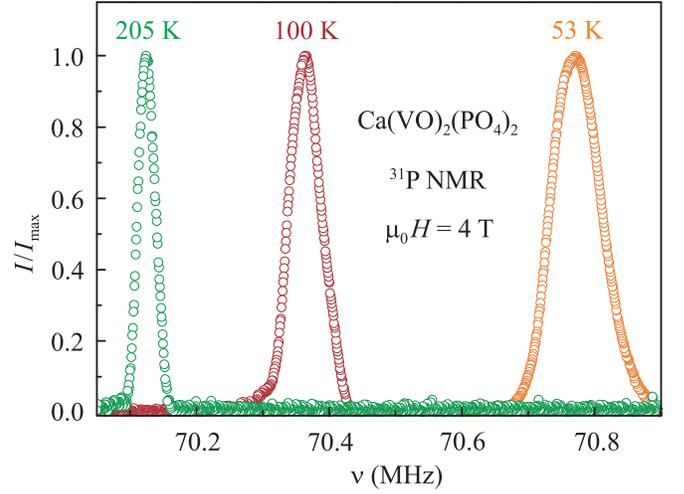}
\caption{\label{fig_nmr}(Color online) $^{31}$P NMR spectra for
Ca(VO)$_{2}$(PO$_{4}$)$_{2}$ measured at 4~T by the Fourier transform of the
spin-echo signal above $T_{N}$. The representative spectra show the
line shift with temperature. All the spectra are normalized to unity
by dividing the intensity ($I$) by the maximum intensity
($I_{\max}$).}
\end{figure}

For both the compounds, the $^{31}$P NMR spectra consist
of a single and narrow spectral line as is expected for $I=1/2$
nuclei.\cite{nath2005, nath2008a} The single spectral line  implies
that both Ca(VO)$_{2}$(PO$_{4}$)$_{2}$ and
Sr(VO)$_{2}$(PO$_{4}$)$_{2}$ have a unique $^{31}$P site consistent with the structural data. Additionally, narrow line is also a signature of good sample quality. Figure~\ref{fig_nmr} shows the
representative spectra for Ca(VO)$_{2}$(PO$_{4}$)$_{2}$ above the transition temperature. The line width and the line shift were found to be temperature dependent.

\begin{figure}
\includegraphics{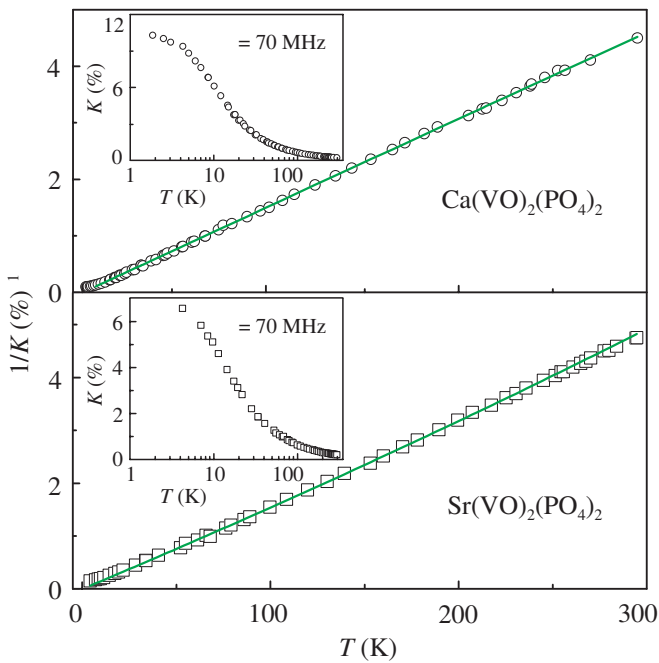}
\caption{\label{fig_shift}
(Color online) Inverse of $^{31}$P NMR shift ($1/K$) vs. $T$ for Ca(VO)$_{2}$(PO$_{4}$)$_{2}$ (upper panel) and Sr(VO)$_{2}$(PO$_{4}$)$_{2}$ (bottom panel). Circles and squares indicate the experimental data, while the solid lines show the Curie-Weiss fits. The insets present the $K$ vs. $T$ curves.}
\end{figure}

The temperature dependence of the NMR shift $K(T)$
(Fig.~\ref{fig_shift}) behaves similar to the bulk susceptibility
$\chi$ (see Fig.~\ref{fig_chi}). NMR has an important advantage over
$\chi$ for the determination of magnetic parameters. At low
temperatures, extrinsic Curie-type paramagnetic contribution often
affects the bulk susceptibility, while in case of NMR this
contribution broadens the spectral line but does not contribute to
the line shift. Therefore, it is sometimes more reliable to extract
the magnetic parameters from the temperature dependence of the NMR
shift rather than from the bulk susceptibility.

\begin{figure}
\includegraphics{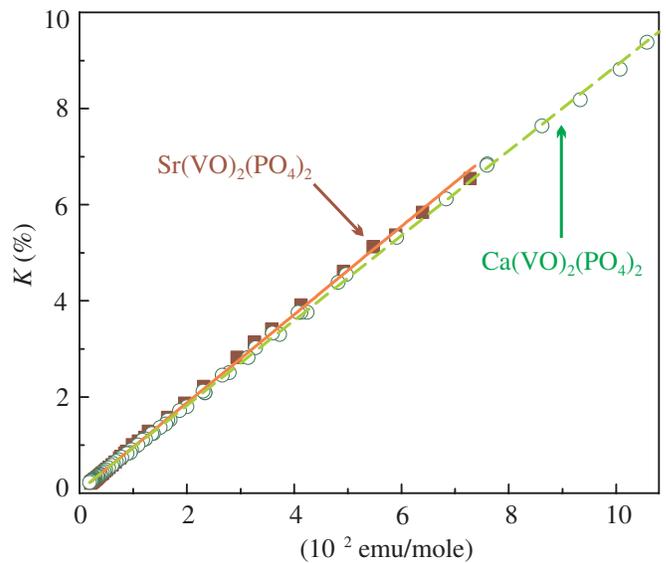}
\caption{\label{fig_shift-chi} (Color online) $^{31}$P NMR shift ($K$) vs. bulk susceptibility ($\chi$) with temperature as an implicit parameter. We employ the $\chi$ data collected at the applied field of 4~T, since this field was used for the NMR measurements. The solid and dashed lines show the linear fits with Eq. (\ref{Shift}).}
\end{figure}

The hyperfine Hamiltonian for $^{31}$P can be written in the form
$\hat H=-\gamma\hbar\,\Iv A_{hf}\Sv$, with $\Iv$ and $\Sv$ being
dimensionless nuclear and electron spins, respectively. The NMR
shift $K$ is a direct measure of the uniform spin susceptibility
$\chi_{\text{spin}}$. Quite generally, their relation is written as
follows:
\begin{equation}
K=\delta + \left(  \frac{A_{hf}}{N_{A}\mu_{B}}\right)  \chi_{\text{spin}}\label{Shift}%
\end{equation}
where $\delta $  is the temperature-independent chemical shift,
which is almost negligible, $N_{A}$ is the Avogadro constant, and
$A_{hf}$ is the transferred hyperfine coupling between the electrons
and the probing nuclei. Therefore to calculate $A_{hf}$, one should
use $K$ vs. $\chi_{\text{spin}}$ plot with $T$ as an implicit
parameter. Since above 3 K the extrinsic paramagnetic contribution
is negligible for both the compounds, we use the bulk susceptibility
$\chi$ instead of the spin susceptibility $\chi_{\text{spin}}$. The
resulting plots are shown in Fig.~\ref{fig_shift-chi}. The fits yield $A_{hf}$ = ($4920\pm 200$) and ($5125\pm 200$) Oe/$\mu_{B}$ for Ca(VO)$_{2}$(PO$_{4}$)$_{2}$ and
Sr(VO)$_{2}$(PO$_{4}$)$_{2}$, respectively. The linearity of the $K$
vs. $\chi_{\text{spin}}$ plots confirms that by measuring $K(T)$ we
can trace $\chi_{\text{spin}}(T)$ properly. The obtained hyperfine
couplings for $^{31}$P are of the same order as the values previously reported
for vanadium phosphates\cite{furukawa1996,kikuchi1999,kikuchi2001}
and indicate a sizeable hybridization of P and V orbitals mediated
by $2p$ orbitals of oxygen.

To extract the magnetic parameters, we fitted $K(T)$ curves above 15
K with the Curie-Weiss law $\chi=C/(T+\theta_{\CW})$. The
resulting values of the effective moment ($\mu_{\eff}^K$) and the
Curie-Weiss temperature ($\theta_{\CW}^K$) are listed in
Table~\ref{Chi_parameters}. These values are in good agreement with
that obtained from the analysis of the bulk susceptibility.

\begin{figure}
\includegraphics{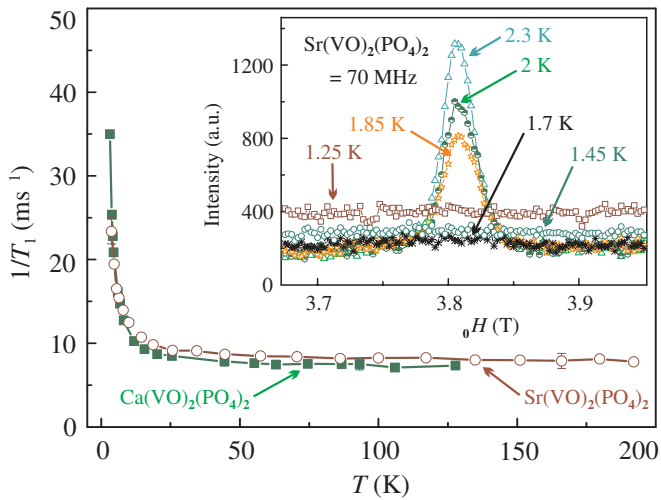}
\caption{\label{fig_t1}(Color online) Spin-lattice relaxation rate ($1/T_{1}$) vs. temperature ($T$) for the M(VO)$_2$(PO$_4)_2$ compounds.
The inset shows the field sweep spectra of Sr(VO)$_2$(PO$_4)_2$
measured at the transmitter frequency of $70$ MHz and temperatures
close to $T_{N}$.}
\end{figure}

For the $1/T_{1}$ experiment, the frequencies of the central positions of the corresponding
spectra at $70$ MHz have been excited. For a spin-1/2 nucleus, the
longitudinal magnetization recovery is expected to follow a single
exponential behavior. In M(VO)$_{2}$(PO$_{4}$)$_{2}$, the recovery
of the nuclear magnetization after an inverting pulse can indeed be
described by a single exponential,
$\frac{1}{2}\left(\frac{M(\infty)-M(t)}{M(\infty)}\right)=A_{1}\exp\left(-t/T_1\right)+C$,
where $M(t)$ is the nuclear magnetization at a time $t$ after an
inverting pulse, while $A_1$ and $C$ are time-independent constants.
Temperature dependences of $1/T_{1}$ are presented in
Fig.~\ref{fig_t1}. The spin-lattice relaxation rates are temperature-independent above 20 K and rapidly increase with the reduction in
temperature below 20 K. The temperature-independent behavior of
$1/T_{1}$ is typical for the paramagnetic regime with fast and
random fluctuations of electronic spins.\cite{moriya1956} The
critical divergence in the vicinity of $T_{N}$ corresponds to the
slowing down of fluctuating moments and indicates the approach to
the state with long-range magnetic ordering.

Furthermore, the $1/T_1$ data enable to estimate exchange couplings
in the system under investigation. According to
Moriya,\cite{moriya1956} the high-temperature limit of $1/T_1$ for a
system of localized moments can be expressed as follows:
\begin{equation}
\left.\left(\dfrac{1}{T_1}\right)\right|_{T\rightarrow\infty}=\frac{\gamma^{2}}{2}\frac{S(S+1)}{3}\frac{\sqrt{2\pi}}{\omega_{\text{E}}}\times \sum\limits_{k,i,j}|A_{ij}^{k}|^{2}
\label{T1}
\end{equation}
where $A_{ij}^{k}$ ($i,j=x,y,z$) are the components of the hyperfine
tensor due to the $k^{\text{th}}$ magnetic atom. The Heisenberg exchange frequency
$\omega_{\text{E}}$ is defined as
$\omega_{\text{E}}=J_{c}(k_B/\hbar)\sqrt{2zS(S+1)/3}$,
$J_c=\sqrt{\sum_i J_i^2}$ is the thermodynamic energy scale of the
exchange couplings, and $z=z_i=2$ (as introduced in Sec.~\ref{heat}).\cite{melzi2001} Using the relevant parameters and the experimental high-temperature relaxation rates $8$ and $8.5$ ms$^{-1}$ (see Fig.~\ref{fig_t1}), we find $J_{c}^{\,T_1}\simeq 13.4$ and $14.1$ K for Ca(VO)$_{2}$(PO$_{4}$)$_{2}$ and Sr(VO)$_{2}$(PO$_{4}$)$_{2}$, respectively. These values are in reasonable agreement with the estimates $J_c^{\,C}\simeq 10$ K from the specific heat (see Sec.~\ref{heat}).

In the spectral measurements at low temperatures (slightly above
$T_N$), a broad background signal appears along with the central
peak for both the compounds. At $T_{N}$, the central peak vanishes,
and the broad background signal becomes prominent extending over a
large field range of about $5$ T. This effect is illustrated in
the inset of Fig.~\ref{fig_t1} for Sr(VO)$_2$(PO$_4)_2$. The central
line disappears at 1.7 K consistent with the ordering temperature of
1.8~K as determined by the specific heat measurements (see the inset
of Fig.~\ref{fig_cph} and note that we use the $T_N$ value at 4~T,
since the NMR measurements are carried out at this applied field).
The broadening of our spectra upon approaching $T_N$ from above is
the usual behavior expected for the magnetic ordering. Below $T_{N}$,
the spectral intensity melts into a broad background signal which
gives strong evidence of a large distribution of internal static
fields in the ordered state. Similar broadening of the spectral line
has been observed for many other materials in the magnetically
ordered state (see, e.g., Refs. \onlinecite{gavilano2000} and~\onlinecite{vonlanthen2002}).

\section{Band structure}
\label{band}
The M(VO)$_2$(PO$_4)_2$ compounds reveal quite complicated spin systems with numerous superexchange paths; therefore the presented experimental data do not allow to determine the individual exchange couplings unambiguously. To overcome this difficulty, we turn to band structure calculations, estimate individual exchange couplings and construct a microscopic model of the exchange interactions. As we will show below, the computational analysis of the M(VO)$_2$(PO$_4)_2$ compounds is also rather difficult, and one can hardly expect the reliable quantitative estimates of all the exchange couplings in the systems under investigation. Nevertheless, we succeed to establish a reasonable spin model and provide a plausible explanation of the frustration (Sec.~\ref{discussion}).

Experimental data show the similarity of the magnetic properties of the M(VO)$_2$(PO$_4)_2$ compounds with M = Ca and Sr. We calculated band structures for both the compounds, analyzed exchange couplings and did not find any considerable differences between the two systems. Therefore, in the following we will discuss the calcium compound only. The whole discussion is applicable to Sr(VO)$_2$(PO$_4)_2$ as well.

\subsection{LDA and tight-binding model}
\begin{figure}
\includegraphics{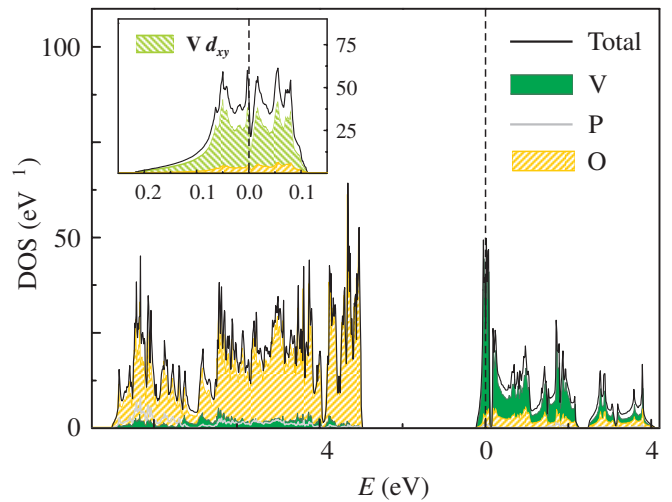}
\caption{\label{fig_dos}
(Color online) LDA density of states for Ca(VO)$_{2}$(PO$_{4}$)$_{2}$. The Fermi level is
at zero energy. The calcium contribution is negligible in the whole energy range, therefore it is not presented in the figure. The inset zooms the image near the Fermi level, only V $3d_{xy}$ and oxygen contributions are shown.}
\end{figure}

The LDA density of states plot for Ca(VO)$_2$(PO$_4)_2$ is shown in
Fig.~\ref{fig_dos}. Valence bands below $-3$ eV are mainly formed by
oxygen orbitals, while the states near the Fermi level have
predominantly vanadium character with an admixture of oxygen. The
phosphorous and calcium contributions to these states are tiny and
hardly visible in the figure. Note that the energy spectrum is
gapless in evident contradiction with the green color of
Ca(VO)$_2$(PO$_4)_2$. This is a typical failure of LDA due to an
underestimate of strong electron-electron correlations in the V $3d$
shell. Local spin density approximation (LSDA)+$U$ calculations
readily reproduce the insulating spectrum with an energy gap of
$2.5-2.8$ eV. The latter values are in reasonable agreement with the
sample color.
\begin{table}
\caption{The hopping parameters $t_i$ of the tight-binding model and the resulting antiferromagnetic couplings $J_i^{\AFM}$ for Ca(VO)$_{2}$(PO$_{4}$)$_{2}$. Only the interactions involving PO$_4$ tetrahedra are listed. The effective on-site Coulomb repulsion potential $U_{\eff}=4.5$ eV.}%
\label{table_tb}%
\begin{ruledtabular}
\begin{tabular}{cccccc}
& $t_1$ & $t_2$ & $t_3$ & $t_4$ & $t_5$ \\
$t$ (meV) & $-12$ & $-33$ & $36$ & $-3$ & $-3$ \\\hline
& $J_1$ & $J_2$ & $J_3$ & $J_4$ & $J_5$ \\
$J^{\text{AFM}}$ (K) & $1.5$ & $11.3$ & $13.4$ & $0.1$ & $0.1$ \\
\end{tabular}
\end{ruledtabular}
\end{table}

\begin{figure}[b]
\includegraphics{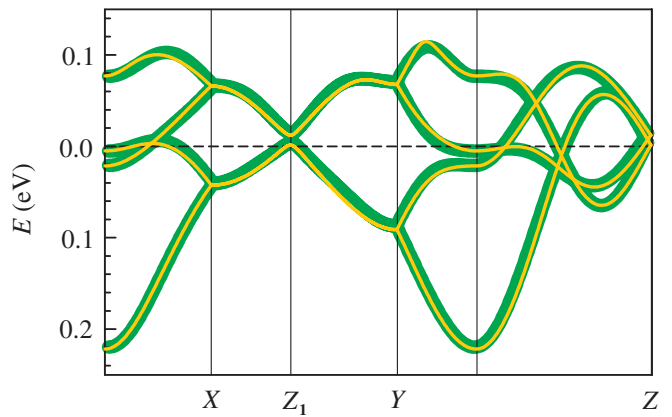}
\caption{\label{fig_bands}
(Color online) LDA band structure of Ca(VO)$_{2}$(PO$_{4}$)$_{2}$ near the Fermi level (thin light lines) and the fit with the tight-binding model (thick darker lines). The Fermi level is at zero energy. The notation of $k$ points is as follows: $\Gamma(0,0,0)$, $X(0.5,0,0)$, $Z_1(0.5,0.5,0)$, $Y(0,0.5,0)$, and $Z(0,0,0.5)$ (the coordinates are given along $k_x$, $k_y$, and $k_z$ in units of the respective reciprocal lattice parameters $4\pi/a$, $4\pi/b$, and $4\pi/c$). The points $Z$ and $Z_1$ are equivalent due to the face-centered symmetry of the unit cell.
}
\end{figure}

To study exchange interactions, we construct an effective model that
includes the relevant, half-filled bands only. Thus, we focus on the
bands that are close to the Fermi level and have predominantly
vanadium character. According to the discussion in
Sec.~\ref{structure}, distorted octahedral coordination of vanadium
gives rise to a non-degenerate $d_{xy}$ ground state. Indeed, we
find four bands formed by V $d_{xy}$ orbitals in the energy range
between $-0.2$ and $0.1$ eV (see the inset of Fig.~\ref{fig_dos} and
Fig.~\ref{fig_bands}). These bands correspond to four vanadium atoms
in the primitive cell of Ca(VO)$_{2}$(PO$_{4}$)$_{2}$ and are used
for a tight-binding fit of the transfer integrals (hoppings)
relevant for the AFM exchange interactions. The resulting transfer
integrals ($t$) are introduced to the extended Hubbard model, and
the correlation effects are taken into account explicitly via an
effective on-site repulsion potential $U_{\eff}$.\cite{foot1} In our
case $t\ll U_{\eff}$, hence the Hubbard model at half-filling can be
reduced to a Heisenberg model for the low-lying (i.e., spin)
excitations. Thus, we are able to estimate AFM contributions to the
exchange couplings as $J_i^{\AFM}=4t_i^2/U_{\eff}$. Similar to
Ref.~\onlinecite{tsirlin2008}, we use $U_{\eff}=4.5$~eV -- a
representative value for vanadium oxides.

An advantage of the TB approach is the possibility to estimate all
the exchange couplings in the system under investigation. Yet the TB
fit is sometimes non-unique, and this is the case for
Ca(VO)$_2$(PO$_4)_2$ due to the presence of numerous NN and NNN
hoppings. To get an unambiguous solution, we assume that the leading
interactions run via PO$_4$ tetrahedra (i.e., correspond to
$t_1-t_5$ introduced in Sec.~\ref{structure}). This assumption looks
reasonable, since magnetic interactions in transition metal
phosphates are usually mediated by PO$_4$ groups (see, e.g.,
Refs.~\onlinecite{tsirlin2008}, \onlinecite{kini2006}, \onlinecite{nath2008a}, and
\onlinecite{johannes2006}). Moreover, phosphorous gives larger
contribution to the states near the Fermi level as compared to
calcium. Thus, the V--O--P--O--V superexchange paths should be
favorable.

The resulting TB fit is in perfect agreement with the LDA band
structure (see Fig.~\ref{fig_bands}). Table~\ref{table_tb} lists the
hoppings involving PO$_4$ tetrahedra and the respective exchange
integrals.\cite{foot8} The strongest AFM coupling runs between
non-parallel structural chains ($J_3$), the other strong coupling
occurs between parallel chains ($J_2$), and all the other AFM
couplings are weaker at least by a factor of eight. The magnitude of
$J's$ is consistent with the experimental data, although the precise
values are somewhat overestimated. For example, considering $J_2$
and $J_3$ only, we find $J_c=17.7$ K that slightly exceeds the
$J_c^{\,T_1}\approx 13$ K estimate from NMR and is well above the estimate $J_c^{\,C}\approx 10$ K from the specific heat.

\subsection{LSDA+$U$, results}
\label{lsda+u-results}
The spin system formed by $J_2$ and $J_3$ is 3D and non-frustrated. To get a frustrated scenario, one has to consider additionally FM interactions using LSDA+$U$ calculations. The regular approach deals with the construction of a supercell enabling different patterns of spin ordering, the mapping of the resulting energies onto a Heisenberg model, and the estimate of individual exchange integrals (see, e.g., Ref.~\onlinecite{tsirlin2008}). However, this approach seems to be inappropriate for Ca(VO)$_2$(PO$_4)_2$, as the magnetic interactions are very weak, and the change in the total energy due to the variation of spin ordering is of the order of 10 K (i.e., $10^{-5}$ Hartree or $\sim 10^{-9}$ of the total energy). Thus, we have to use the smallest unit cell in order to achieve the highest possible accuracy, and only sums of the exchange couplings can be estimated. Nevertheless, one may expect that the TB results on $J_i^{\AFM}$ will enable to resolve FM interactions (at least, qualitatively).

The results of the LSDA+$U$ calculations are listed in
Table~\ref{table_lsda+u}. The most stable spin configuration is
formed by ferromagnetic structural chains that are coupled both
ferromagnetically (for parallel chains) and antiferromagnetically
(for non-parallel chains). We find the same ground state for
$U_d=4-6$ eV, although the exchange couplings strongly depend on the
$U_d$ value. This problem will be addressed below
(Sec.~\ref{lsda+u-discussion}) after a comparison of the LSDA+$U$
results with the experimental data and the TB estimates.

\begin{table}[ptb]
\caption{LSDA+$U$ estimates for the exchange integrals in
Ca(VO)$_{2}$(PO$_{4}$)$_{2}$ and the resulting Curie-Weiss
temperatures $\theta_{\CW}^{\protect\raisebox{1.5pt}{\scriptsize\,\text{calc}}}$.}
\label{table_lsda+u}%
\begin{ruledtabular}
\begin{tabular}{cccccc}
$U_d$ (eV) & $J_1+J_2$ (K) & $J_3$ (K) & $J_4+J_5$ (K) & $\theta_{\CW}^{\raisebox{1.5pt}{\scriptsize\,\text{calc}}}$ (K) \\
4 & $2.9$ & $4.4$ & $6.0$ & $6.7$ \\
5 & $-14.0$ & $4.1$ & $9.2$ & $-0.4$ \\
6 & $-40.5$ & $3.7$ & $14.0$ & $-11.4$ \\
\end{tabular}
\end{ruledtabular}
\end{table}
As we have mentioned in Sec.~\ref{method}, $U_d$ is an adjustable parameter that depends on the basis set, crystal structure of the compound under investigation, and the objective parameters of the calculation. Usually, one has to fit $U_d$ to some observable quantity (energy gap, magnetic moments, etc.). It is preferable to use the quantity that is related to the objective parameters, since different objective parameters may require the application of different $U_d$ values. Below, we will employ the Curie-Weiss temperature $\theta_{\CW}$ for this purpose. The Curie-Weiss temperature is the second-order term in the high-temperature series expansion of the magnetic susceptibility. According to Ref.~\onlinecite{johnston2000}, for spin-1/2 system $\theta_{\CW}=1/4\sum_iz_iJ_i=1/2\sum_iJ_i$. Thus, $\theta_{\CW}$ is directly related to the exchange couplings and provides reasonable justification for the choice of the $U_d$ value.

Clearly, the results for $U_d=5$ eV are in good agreement with the experimental estimate $\theta_{\CW}\simeq -0.3$ K, while for $U_d=4$ and 6 eV the compliance of the experimental and computational results is quite poor (see Table~\ref{table_lsda+u}). Thus, the optimal value of $U_d$ for Ca(VO)$_2$(PO$_4)_2$ is 5 eV.\cite{foot9} Yet in contrast to other spin-1/2 systems,\cite{drechsler2007,tsirlin2008,johannes2006} the slight variation of $U_d$ results in a huge change in the exchange couplings (see Table~\ref{table_lsda+u}). This effect will be further discussed in Sec.~\ref{lsda+u-discussion}. In the rest of this subsection, we will focus on the exchange couplings calculated with $U_d=5$ eV.

According to Tables~\ref{table_tb} and \ref{table_lsda+u},
$J_1+J_2\ll J_1^{\AFM}+J_2^{\AFM}$ indicating considerable FM
contribution to either $J_1$ and/or $J_2$. $J_3$ is smaller than
$J_3^{\AFM}$ and positive, hence there is also a FM contribution to
$J_3$, but the overall interaction remains AFM. Finally, $J_4+J_5\gg
J_4^{\AFM}+J_5^{\AFM}$, i.e., the LSDA+$U$ and TB estimates for
these couplings are inconsistent. One may think that additional
interactions contribute to the LSDA+$U$ value of $J_4+J_5$, but such
contributions yield $2-3$ K only\cite{foot2} and do not explain the
controversy. Our experience shows that the TB approach is highly reliable
for estimating exchange couplings in spin-1/2
systems.\cite{johannes2006,janson2007,tsirlin2008} In contrast to
LSDA+$U$, the TB approach includes a simple explicit relation
between $J_i$ and $U_{\eff}$, while the underlying LDA calculation
is \textit{ab initio} and does not include any adjustable
parameters.  Therefore, we have to admit the failure of LSDA+$U$ to
provide valid quantitative estimates of the weak exchange couplings
in Ca(VO)$_2$(PO$_4)_2$ (see Sec.~\ref{lsda+u-discussion} for
further discussion). Nevertheless, the LSDA+$U$ results are helpful
to get a qualitative understanding of the system under
investigation.

To get an idea about the FM interactions in Ca(VO)$_2$(PO$_4)_2$, we
consider vanadium--vanadium separations for $J_1-J_3$. The shortest
separation (3.46 \r A) corresponds to $J_1$, while $J_2$ and $J_3$
reveal larger distances of 6.15 and 4.35 \r A, respectively. FM
interactions are short-range; therefore it is natural to suggest the
strongest FM contribution to be that of $J_1$. Since $J_1^{\AFM}$ is
close to zero, the overall interaction $J_1$ should be FM. Unlike
$J_1$, $J_2$ and $J_3$ are AFM, although the FM contributions may
also be non-negligible and reduce the absolute values of the
exchange couplings. The presence of the ferromagnetic in-chain
interaction $J_1$ is consistent with our studies of other vanadium
phosphates having similar chains of corner-sharing
octahedra.\cite{foot3}

Thus, the basic microscopic model for Ca(VO)$_2$(PO$_4)_2$ includes three exchange couplings: ferromagnetic $J_1$ and antiferromagnetic $J_2$ and $J_3$ interactions. The resulting spin system is 3D and frustrated (see Fig.~\ref{fig_frustration} and Sec.~\ref{discussion}). Unfortunately, the computational results do not enable quantitative estimates of individual exchange couplings (see the above discussion and Sec.~\ref{lsda+u-discussion}). Therefore, we do not make further numerical comparisons with the experimental data. Clearly, the presence of both the FM and AFM couplings is qualitatively consistent with the strongly reduced $\theta_{\CW}$. Moreover, band structure calculations suggest rather weak (about 10 K) exchange interactions in Ca(VO)$_2$(PO$_4)_2$ consistent with the experimental energy scale $J_c=10-15$ K. We are convinced that the agreement between the band structure results and the experimental data is reasonable, especially taking into account the complexity of the spin system and the weakness of the exchange couplings.

\subsection{LSDA+$U$: The influence of $U_d$}
\label{lsda+u-discussion}
To understand the $J$ vs. $U_d$ trends, one should consider exchange integrals as composed of the AFM and FM contributions: $J=J^{\AFM}+J^{\FM}$. Within one-band Hubbard model, $J^{\AFM}\sim t^2/U$, while $J^{\FM}$ is independent of $U$. Therefore, the overall $J$ should be reduced with the increase in $U$. The repulsion potential $U_d$ is a parameter of the computational method rather than the true on-site Coulomb repulsion, and $J^{\AFM}$ is not necessarily proportional to $1/U_d$. Yet the increase in $U_d$ tends to improve the localization of $3d$ electrons hence suppressing AFM interactions. Thus, one would expect that antiferromagnetic exchange couplings should be reduced with the increase of $U_d$. In fact, the strongest coupling usually shows $\sim 1/U_d$ behavior, while other exchange integrals are nearly independent of $U_d$ (see, e.g., Refs.~\onlinecite{tsirlin2008} and \onlinecite{tsirlin2008a}). If $J^{\AFM}\gg J^{\FM}$, the overall $J$ is antiferromagnetic for any reasonable value of $U_d$. However, for small $J^{\AFM}\approx J^{\FM}$ the situation will be different, and the resulting $J$ may be positive for small $U_d$ and negative for large $U_d$.

The above considerations naturally explain, why FM interactions manifest themselves (hence reducing $\theta_{\CW}$) at $U_d\geq 5$ eV only. However, the FM interactions are further enhanced with the increase in $U_d$, and this effect has a different origin. LSDA+$U$ treats electronic correlations within mean-field approximation; therefore the filled states of vanadium are merely shifted to lower energies as the repulsive potential $U_d$ is applied. At lower energies, oxygen states dominate; hence vanadium--oxygen hybridization is enhanced with the increase in $U_d$. Indeed, our calculations reveal the decrease in the magnetic moment of vanadium and the partial spin polarization of oxygen atoms as $U_d$ is increased. In case of Ca(VO)$_2$(PO$_4)_2$, this effect is even more pronounced as compared to other vanadium compounds.\cite{tsirlin2008,tsirlin2008a} Moreover, the exchange couplings are very weak, and the overestimated vanadium--oxygen hybridization results in unrealistic results for $J_i$ (note the $J_1+J_2$ value of $-40$ K at $U_d=6$ eV).

In summary, we should point to two complications that arise during LSDA+$U$ calculations for vanadium oxides with very weak magnetic interactions. First, one has to apply a relatively large $U_d$ in order to suppress AFM couplings and to reveal FM contributions. Second, large $U_d$ tends to overestimate vanadium--oxygen hybridization and to produce unrealistic enhancement of the exchange couplings. There is an optimal $U_d$ value [5 eV in case of Ca(VO)$_2$(PO$_4)_2$] that gives rise to a reasonable solution. However, even at this value both problems are retained. In general, we think that LSDA+$U$ estimates of weak exchange couplings in highly complex structures are not accurate enough to provide quantitatively correct results. Therefore, we mainly rely on the TB model in our analysis. Nevertheless, LSDA+$U$ estimates facilitate a qualitative understanding of the coupling scenario resulting in a reasonable microscopic description of the magnetic behavior.

\section{Discussion}
\label{discussion} Our experimental results show that the complex
vanadium phosphates M(VO)$_2$(PO$_4)_2$ are strongly frustrated spin
systems with competing FM and AFM exchange couplings. Magnetic
susceptibility data  reveal a maximum at $T_{\max}^{\chi}\simeq 3$ K
and nearly vanishing Curie-Weiss temperatures $\theta_{\CW}\leq 1$
K. For simple spin systems with a single leading exchange coupling,
$T_{\max}^{\chi}$ and $\theta_{\CW}$ are usually close to each
other, while the $\theta_{\CW}\ll T_{\max}^{\chi}$ regime implies
the presence of several different magnetic interactions in the
system under investigation. As we have stated above, $\theta_{\CW}$
is equal to half of the sum of all the exchange integrals. Thus, the
low $\theta_{\CW}$ value points to the presence of both FM and AFM
interactions that compensate each other. Further on, the low N\'eel
temperatures ($T_N\approx T_{\max}^{\chi}/2$) indicate that
long-range ordering in the system is impeded by either low
dimensionality and/or magnetic frustration.

Low-temperature specific heat provides the information on the spin excitations and presents one of the substantial characteristics of quantum magnets. In systems with strong quantum fluctuations, the specific heat maximum is reduced due to the suppression of correlated spin excitations. The maximum of the magnetic specific heat of the M(VO)$_2$(PO$_4)_2$ compounds ($C_{\mg}^{\max}$) equals to $0.25R$ only. This value is comparable to that for the spin-1/2 triangular lattice, a system with strong geometrical frustration, and lies well below $C_{\mg}^{\max}$ typical for non-frustrated low-dimensional spin systems. Thus, strong quantum fluctuations are present in M(VO)$_2$(PO$_4)_2$, and these fluctuations are likely caused by the magnetic frustration.
\begin{figure}
\includegraphics{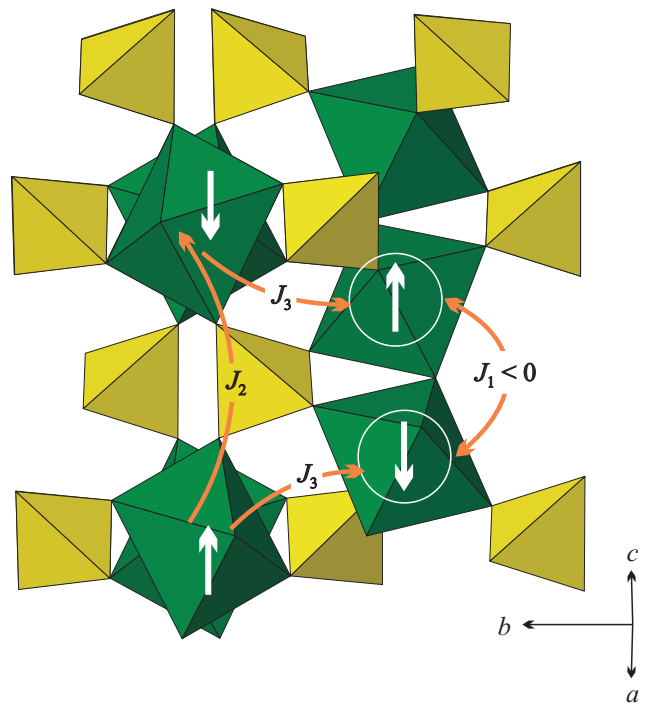}
\caption{\label{fig_frustration}
(Color online) Enlargement of the crystal structure of Ca(VO)$_2$(PO$_4)_2$ (see Fig.~\ref{fig_struct}) that shows the proposed scenario of the frustration. Antiferromagnetic inter-chain couplings $J_2$, $J_3$ favor antiparallel orientation of spins for neighboring atoms in the chain (marked by large circles), while the in-chain coupling ($J_1$) is ferromagnetic and would favor parallel alignment of the respective spins.}
\end{figure}

To reveal the origin of the frustration in the M(VO)$_2$(PO$_4)_2$
compounds, we have performed band structure calculations and
estimated the individual exchange couplings. Despite the problems
discussed in Sec.~\ref{lsda+u-discussion}, we succeeded to construct
a reasonable microscopic model that provides a plausible scenario of
the frustration. The coupling within the structural chains (the
chains of corner-sharing VO$_6$ octahedra, see
Fig.~\ref{fig_struct}) is FM ($J_1$), while interchain couplings are
AFM ($J_2$ and $J_3$). The interactions form a quadrangle with one FM
and three AFM sides (Fig.~\ref{fig_frustration}). The FM in-chain
interaction favors the parallel alignment of the spins within the
chain, while the inter-chain AFM interactions favor the antiparallel
alignment. Thus, the spin system is frustrated.

Below, we will use several experimental quantities
($\theta_{\CW},J_c$, and $H_s$) in order to estimate the individual
exchange couplings of the proposed model. Assuming the ground state
revealed by the LSDA+$U$ calculations (see Sec.~\ref{lsda+u-results}),
we find $H_s=2J_3g\mu_B/k_B$ (the Heisenberg model is treated in a
classical way). Using the averaged estimates of
$J_c=(J_c^{\, C}+J_c^{\,T_1})/2$ and
$\theta_{\CW}=(\theta_{\CW}^{\chi}+\theta_{\CW}^K)/2$ and assuming
negative $J_1$, one arrives at $J_1=-9.6$ K, $J_2=3.5$ K, and $J_3=5.4$
K for Ca(VO)$_2$(PO$_4)_2$ and $J_1=-8.9$ K, $J_2=1.8$ K, and $J_3=7.7$~K for Sr(VO)$_2$(PO$_4)_2$. These values should be considered as
rough estimates, since our model includes three interactions only,
and the results of different methods for $J_c$ and $\theta_{\CW}$
are simply averaged. The estimates emphasize the similarity of the
calcium and strontium compounds, although $J_2$ and $J_3$ values are
somewhat different for M = Ca and Sr. The latter result may be
relevant for the different breadth of the specific heat maxima in
the two compounds (see Fig.~\ref{fig_cp-comparison}).

We should emphasize that at least two couplings ($J_1$ and $J_3$) in the M(VO)$_2$(PO$_4)_2$ compounds have similar magnitude, and the spin system is essentially 3D rather than 1D with frustrated inter-chain couplings. The frustration is controlled by the magnitudes of the competing exchange interactions. For example, the decrease of the absolute value of $J_1$ will reduce the frustration, while for AFM $J_1$ the system is non-frustrated at all. To the best of our knowledge, no theoretical results for this model are available. Basically, one may expect the presence of strongly frustrated regions [one is revealed by the M(VO)$_2$(PO$_4)_2$ compounds] and quantum critical points in the respective phase diagram. Further studies of the model and the appropriate materials are desirable.

Finally, we will focus on the structural aspects of the present
study. The main structural feature of the M(VO)$_2$(PO$_4)_2$
compounds deals with the chains of corner-sharing VO$_6$ octahedra.
Such chains are typical for a wide variety of vanadium
oxides.\cite{boudin2000} Within a very straight-forward and naive
approach, one can identify the chains of vanadium polyhedra as spin
chains and arrive at 1D spin system with weak inter-chain couplings.
According to Sec.~\ref{structure}, such an approach is inconsistent
with the orbital state of vanadium, and the actual spin system may
have any dimensionality depending on the exchange couplings via
PO$_4$ tetrahedra or other side groups. In the case of
M(VO)$_2$(PO$_4)_2$, the actual spin system is 3D. Yet in
Sr$_2$VO(VO$_4)_2$ the spin system is 1D (consistent with the naive
expectations), but the spin chains are perpendicular to structural
ones (in contrast to the naive expectations).\cite{kaul2003} In
general, the structures formed by VO$_6$ octahedra and non-magnetic
tetrahedral groups (PO$_4$, V$^{+5}$O$_4$, SiO$_4$, etc.) provide a
promising way for studying different spin systems, and further
investigations could be interesting. However, we should point to the importance of careful thermodynamic measurements and
electronic-structure-based microscopic modeling for proper
understanding of the respective systems. Thus, the interesting
physics of the M(VO)$_2$(PO$_4)_2$ compounds is well hidden behind
the 1D features of the crystal structure and weak magnetic
interactions resulting in the paramagnetic behavior above $4-5$ K.
Only the combination of experimental and computational approaches
enables us to unravel the relevant microscopic mechanisms and to achieve
an insight of the underlying physics.

In conclusion, we have shown that the M(VO)$_2$(PO$_4)_2$ compounds reveal strongly frustrated 3D spin systems with competing FM and AFM interactions. The presence of both FM and AFM interactions is indicated by the vanishing Curie-Weiss temperature $\theta_{\CW}$, while the strong frustration is evidenced by the reduced maximum of the magnetic specific heat. The thermodynamic energy scale of the exchange couplings is $J_c=10-15$~K as shown by the specific heat and NMR data. Band structure calculations suggest a plausible scenario of the frustration with the FM interaction along the structural chains and the AFM interactions between the chains. In the resulting three-dimensional spin system, the frustration is controlled by the magnitudes of the competing exchange couplings. The proposed spin model deserves further experimental and theoretical studies.

\begin{acknowledgments}
The authors are grateful to P. Carretta for his critical suggestions on the NMR results and to Nic Shannon for the fruitful discussions. Financial support of GIF (Grant No. I-811-257.14/03), RFBR (Grant No. 07-03-00890), and the Emmy-Noether program of the DFG as well as the computational facilities of ZIH Dresden are acknowledged.
\end{acknowledgments}


\begin{thebibliography}{99}
\bibitem{foot7}In the strontium compound, the transition anomaly is more pronounced, and the origin of this effect is not clear. Basically, the specific heat anomalies are affected by both intrinsic (dimensionality of the spin system) and extrinsic (sample quality) factors. At present, we can not distinguish which of these factors are relevant for the difference of the transition anomalies for the M(VO)$_2$(PO$_4)_2$ compounds.
\bibitem{foot1}Note that $U_{\eff}\neq U_d$ and these two parameters have fairly different meaning. $U_{\eff}$ is the effective Coulomb repulsion in the mixed V $3d$ -- O $2p$ band, while $U_d$ is the repulsive potential applied to V $3d$ orbitals only.
\bibitem{foot8}The TB model included all the NN and NNN hoppings as well as a number of long-range hoppings. The largest transfer integral not listed in Table~\ref{table_tb} is 10 meV. The respective $J^{\AFM}$ is about 1 K, i.e., well below the strongest interactions $J_2$ and $J_3$. Therefore, in the present discussion we neglect all the exchange couplings that do not run via PO$_4$ tetrahedra.
\bibitem{foot2}The value ``$J_4+J_5$" in Table~\ref{table_lsda+u} sums all the interactions between the atoms in positions $(x,y,z)$ and $(\bar x,\bar y,z)$ or $(x+1/4,\bar y+1/4,z+1/4)$ and $(\bar x+1/4,y+1/4,z+1/4)$ of the $Fdd2$ space group. The TB model indicates two leading hoppings between these atoms: $t_6=-9$ meV and $t_7=-7$~meV. The hoppings provide an overall antiferromagnetic interaction of about 2 K that exceeds $J_4^{\AFM}+J_5^{\AFM}$. However, such an interaction is
still considerably lower than the LSDA+$U$ result.
\end{thebibliography}
\end{document}